\documentclass{PoS}
\pdfoutput=1
\usepackage{graphicx}
\usepackage{amsmath} 
\usepackage{amsfonts}
\usepackage{amssymb}
\usepackage{subfigure}
\usepackage[normalem]{ulem} 
\usepackage{hepunits}
\usepackage{multirow}

\newcommand{\BigO}{\operatorname{O}}
\newcommand{\etal}{\emph{et al.}}

\newcommand{\fig}{Fig.~}
\newcommand{\tab}{Tab.~}
\newcommand{\fm}{{\rm \,fm}}

\title{Topology of dynamical lattice configurations\\
 including results from dynamical overlap fermions}

\ShortTitle{Topology of dynamical lattice configurations}

\author{Falk Bruckmann\\
        Institut f\"ur Theoretische Physik, Universit\"at Regensburg, D-93040 Regensburg, Germany\\
        E-mail: \email{falk.bruckmann@physik.uni-regensburg.de}}
\author{Nigel Cundy\\
        Department of Physics and Astronomy, Seoul National University, Seoul, 151-747, South Korea\\
        E-mail: \email{nigel.cundy@physik.uni-regensburg.de}}
\author{\speaker{Florian Gruber}\\
        Institut f\"ur Theoretische Physik, Universit\"at Regensburg, D-93040 Regensburg, Germany\\
        E-mail: \email{florian.gruber@physik.uni-regensburg.de}}
\author{Thomas Lippert\\
	J\"ulich Supercomputing Centre, Forschungszentrum J\"ulich GmbH, D-52425 J\"ulich, Germany\\
        E-mail: \email{Th.Lippert@fz-juelich.de}}
\author{Andreas Sch\"afer\\
        Institut f\"ur Theoretische Physik, Universit\"at Regensburg, D-93040 Regensburg, Germany\\
        E-mail: \email{andreas.schaefer@physik.uni-regensburg.de}}

\abstract{We investigate how the topological charge density in lattice QCD simulations is affected by violations of chiral symmetry in different fermion actions. To this end we compare lattice configurations generated with a number of different actions including first configurations generated with exact overlap quarks. We visualize the topological profiles after mild smearing. In the topological charge correlator we measure the size of the positive core, which is known to vanish in the continuum limit. To leading order we find the core size to scale linearly with the lattice spacing with the same coefficient for all actions, even including quenched simulations. In the subleading term the different actions vary over a range of about $10$\%. Our findings suggest that non-chiral lattice actions at current lattice spacings do not differ much for this specific observable related to topology, both among themselves and compared to overlap fermions.}

\FullConference{ The XXIX International Symposium on Lattice Field Theory - Lattice 2011\\
July 10-16, 2011\\
Squaw Valley, Lake Tahoe, California}

\begin{document}
\section{Introduction}

Lattice QCD has seen progress towards smaller lattice spacings and realistic pion masses. Chiral symmetry -- a key feature of the continuum theory -- cannot be realized in a straight forward way for doubler free formulations on a lattice. Most of the fermion actions, therefore, have to break chiral symmetry explicitly or they keep only an approximation of it. Since chiral symmetry is an important property of QCD, it is not clear a priori, how harmful 
these approximations are for physics results. The only known way to treat chiral symmetry properly is to use Ginsparg-Wilson fermions, usually realized as overlap fermions, which are numerically very expensive. To avoid uncontrolled effects in state-of-the-art lattice simulations we need benchmark comparisons with results from dynamical overlap fermions especially for observables which are sensitive to chiral symmetry. 

To this end we analyze the topological charge density. Topology is intimately connected to chiral symmetry, both through the chiral anomaly giving mass to the $\eta'$ meson and through the index theorems linking fermionic zero modes to the topological charge. We, therefore, believe that the topological charge density, although a purely gluonic quantity, is well suited to test the effects due to violation of chiral symmetry for different fermion actions \cite{Bruckmann2011b}.

\section{The topological charge density and its correlator}
The topological charge and its (Euclidean) density are defined as
\begin{equation}
\label{eq:qdens}
Q=\int q(x)\,d^4x,\quad
q(x)=\frac{1}{32\pi^2}\, \epsilon_{\mu\nu\rho\sigma}{\rm Tr}\Big(F_{\mu\nu}(x)F_{\rho\sigma}(x)\Big).
\end{equation}
In this work we have focused on the 2-point function of the topological charge density $\langle q(x)q(y)\rangle$
which integrates to the topological susceptibility
\begin{equation}\label{eq:topsuceptibility}
\chi_{\rm top}=\frac{\langle Q^{2}\rangle}{\text{V}}=\int \langle q(0)q(x)\rangle d^{4}x.
\end{equation}

The topological charge, as a genuine continuum quantity, cannot be defined without ambiguities on a lattice.
We use a discretized topological charge density with an improved lattice field strength 
tensor, which combines $1\times1$, $2\times2$ and $3\times3$ loops \cite{Bilson-Thompson2003}.

In order to remove unwanted short range fluctuations, we need filtering methods to extract the relevant long range degrees of freedom. 
The structures which emerge after these filtering procedures are not unique. They strongly depend on parameters like numbers of iterations or mode number. However, in the weak filtering regime one finds that different methods reveal similar local structures if one carefully matches their parameters \cite{Bruckmann2007c,Bruckmann2009a,Bruckmann2010a}.
In this work we use the improved stout smearing algorithm \cite{Moran2008a}.  
While other smearing methods, as well as cooling, destroy topological structures in the long run, this algorithm is designed to preserve instanton-like objects in a wide range of their size parameter. Moreover, 5 steps of this improved stout smearing were found to produce a topological charge density very similar to the fermionic one, $q_{\rm ferm}$ \cite{Ilgenfritz:2008ia}.

\section{Lattice configurations}
\begin{table*}[!t]
\centering
\begin{tabular}{ccccccc}\hline
\textbf{fermion (gauge) action }&$\mathbf{N_{f}}$ & $\mathbf{a[\fm]}$& $\mathbf{V[a^4]}$ & $\mathbf{m_{\pi} [\MeV]}$ &$\mathbf{N_{\rm conf}}$&{\bf Ref.}\\ \hline
twisted mass  (Sym.) &2& $0.10-0.063$ & $24^3 48\, - \,32^3 64$ &$\approx 500 $&20 &\cite{Baron:2009wt}\\ 
twisted mass  (Sym.) &2& $0.10-0.051$ & $20^348\, -\,32^364$ &$ \approx 280 $&20 &\cite{Baron:2009wt}\\  
np. imp. clover (Plaq.) &2& $0.11-0.07$ & $24^3 48$ &$\approx 500$&20& \cite{Gockeler2006}\\  
np. imp. clover (Plaq.) &2& $0.10-0.07$ & $32^3 64$ &$\approx 250 $&20& \cite{Gockeler2006}\\  
asqtad staggered  (LW)&2+1& $0.15-0.09$ & $16^3 48\, - \, 28^{3}96$ &$\approx 500$&5&\cite{Bernard2007a} \\  
top. fixed overlap (Iw.)&2 & $0.12$ & $16^3 32$&$\approx 500 $&20&\cite{Aoki2008} \\  
chirally imp. (LW) &2& $0.15$ &  $16^3 32$ &$\approx 500$& 20 &\cite{Gattringer2009} \\ 
overlap (LW) &2+1& $0.13-0.12$ & $12^3 24$ &$\approx 500$& 15 & \cite{Cundy:2008zc} \\ \hline 
\end{tabular}
\caption{Configurations used in this work. (Abbreviations: Sym. = Symanzik , Plaq. = Plaquette, LW = L\"uscher-Weisz, Iw. = Iwasaki)}\label{tab:configs}
\end{table*}

We compare full dynamical $N_{f}=2$ and $N_{f}=2+1$ flavor configurations from different fermion formulations. They are available through the International Lattice Data Grid \cite{ildg} in a wide range of lattice spacings, lattice volumes and pion masses, see \tab\ref{tab:configs}.

Additionally, we have generated three $N_{\rm f}=2+1$ dynamical overlap ensembles with pion masses, which were measured to be around $550\,\MeV$, at lattice spacings of around $0.12 \fm$ and a lattice volume of $12^324$. For more details on the dynamical overlap configurations and the algorithmic improvements, which made these simulations possible, we refer to Refs.~\cite{Bruckmann2011b,Cundy:2008zc}.
  
\section{Results}
 \begin{figure*}[t!!]
 \centering
  \begin{tabular}{cccc}
 \textbf{dyn. overlap} & \textbf{top. fixed overlap} &\textbf{quenched Iwasaki}& \\
\begin{minipage}[]{0.2\textwidth}
\includegraphics[width=0.9\textwidth]{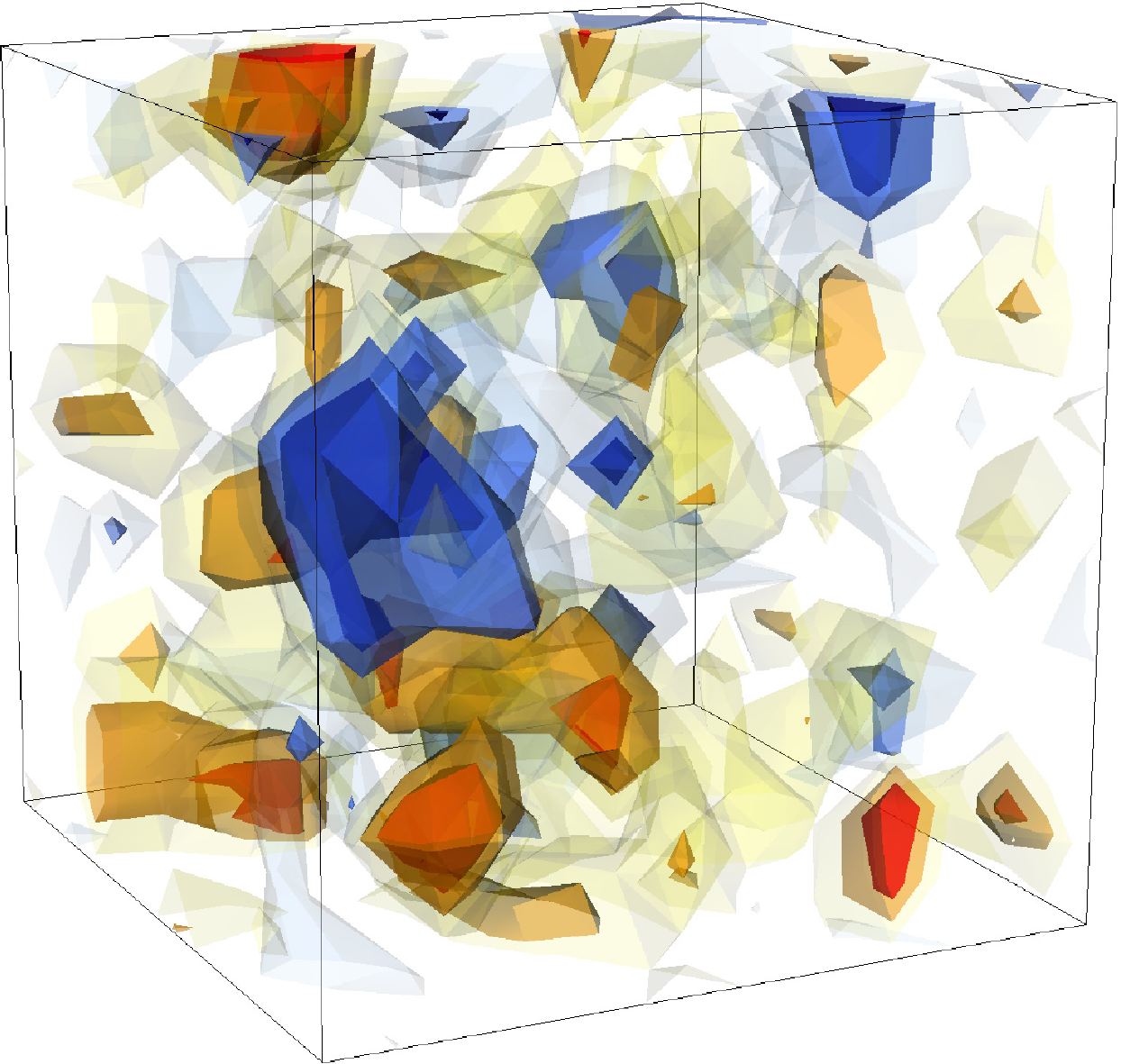}
\end{minipage}  & 
\begin{minipage}[]{0.2\textwidth}
\includegraphics[width=0.9\textwidth]{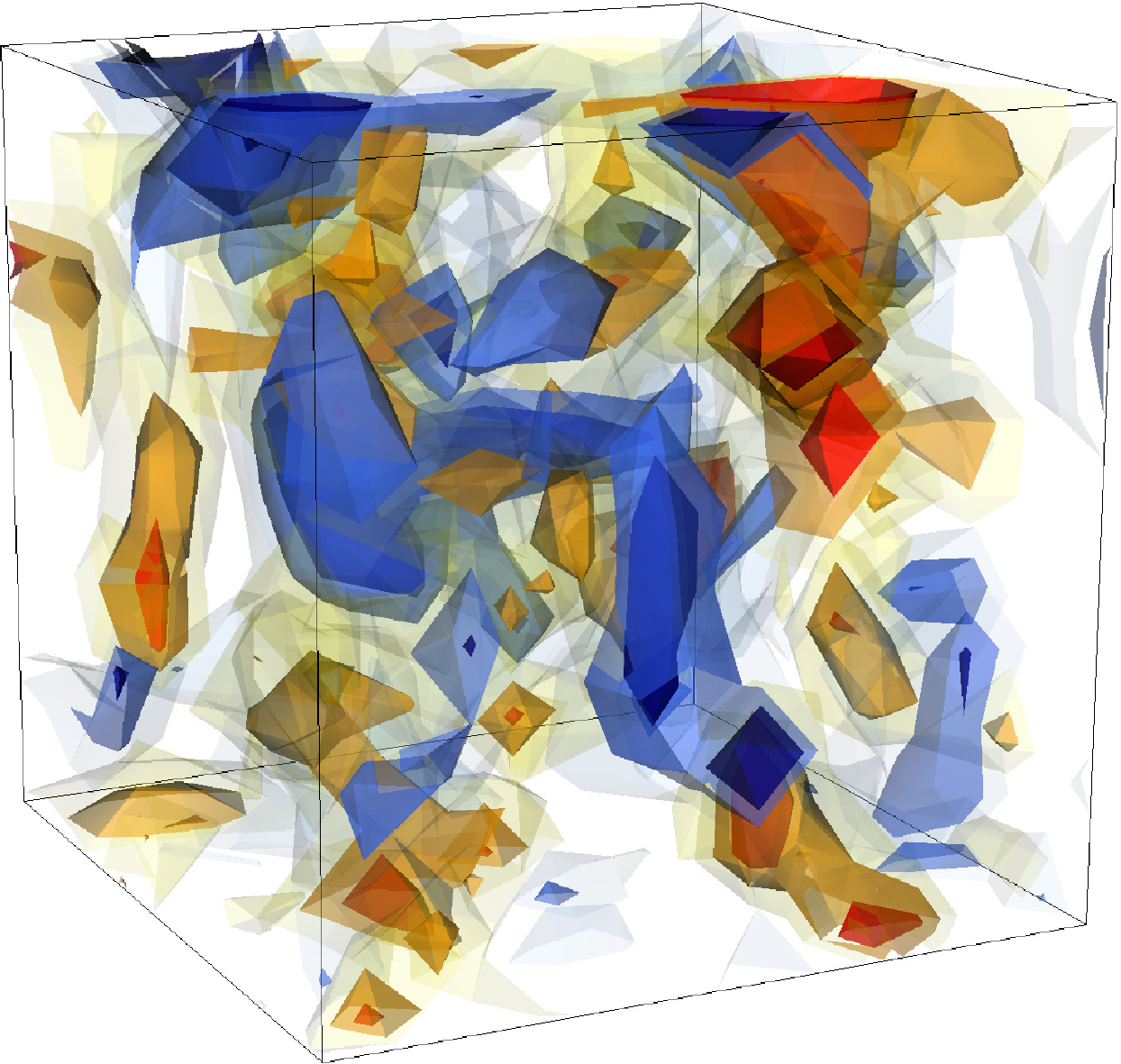}
\end{minipage}  & 
\begin{minipage}[]{0.2\textwidth}
\includegraphics[width=0.9\textwidth]{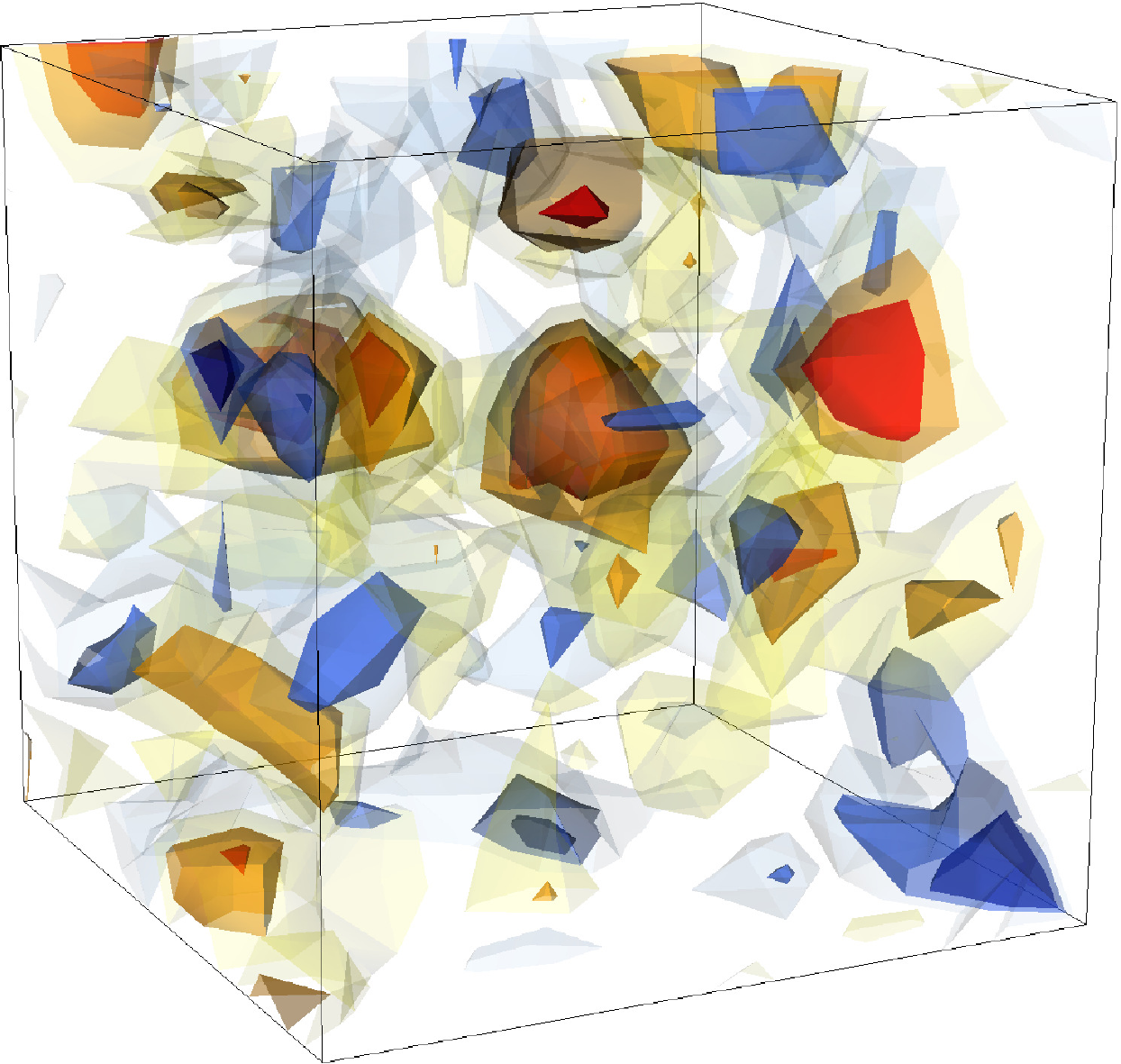}
\end{minipage} &
\\ [24pt]
\textbf{asqtad} & \textbf{clover}& \textbf{quenched plaquette}& \\ 
\begin{minipage}[]{0.2\textwidth}
\includegraphics[width=0.9\textwidth]{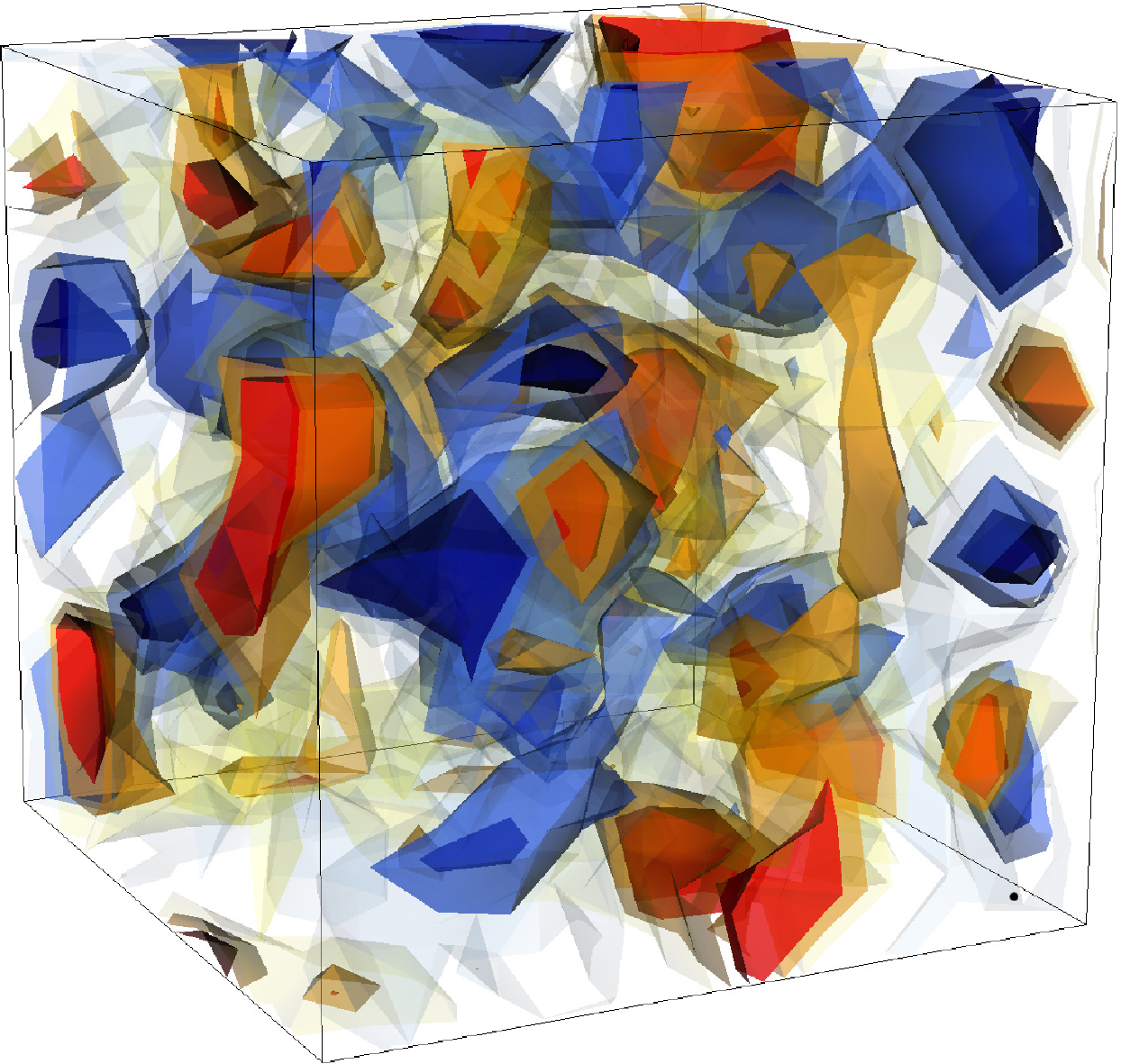}
\end{minipage}  &
\begin{minipage}[]{0.2\textwidth}
\includegraphics[width=0.9\textwidth]{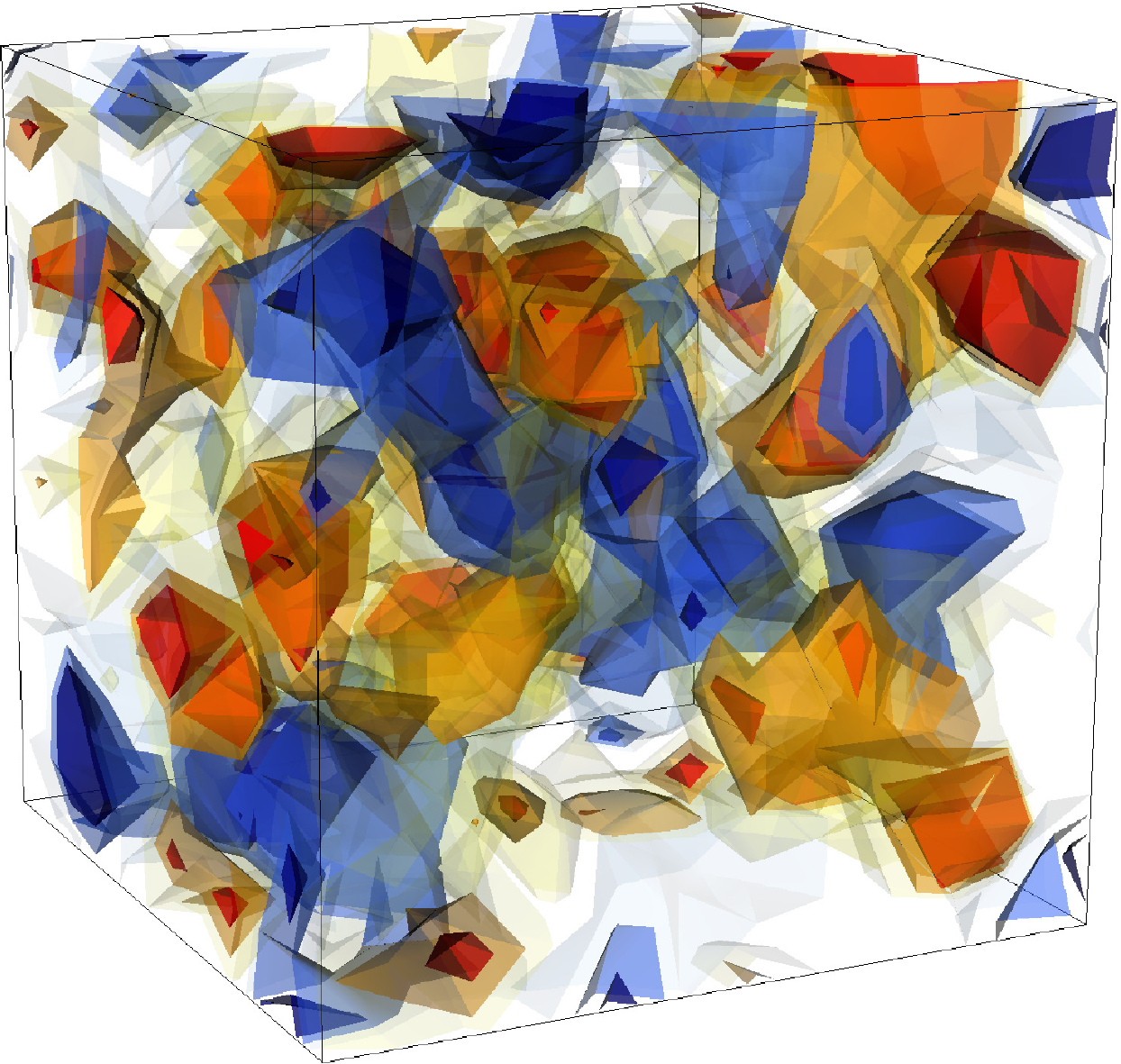}
\end{minipage}& 
\begin{minipage}[]{0.2\textwidth}
\includegraphics[width=0.9\textwidth]{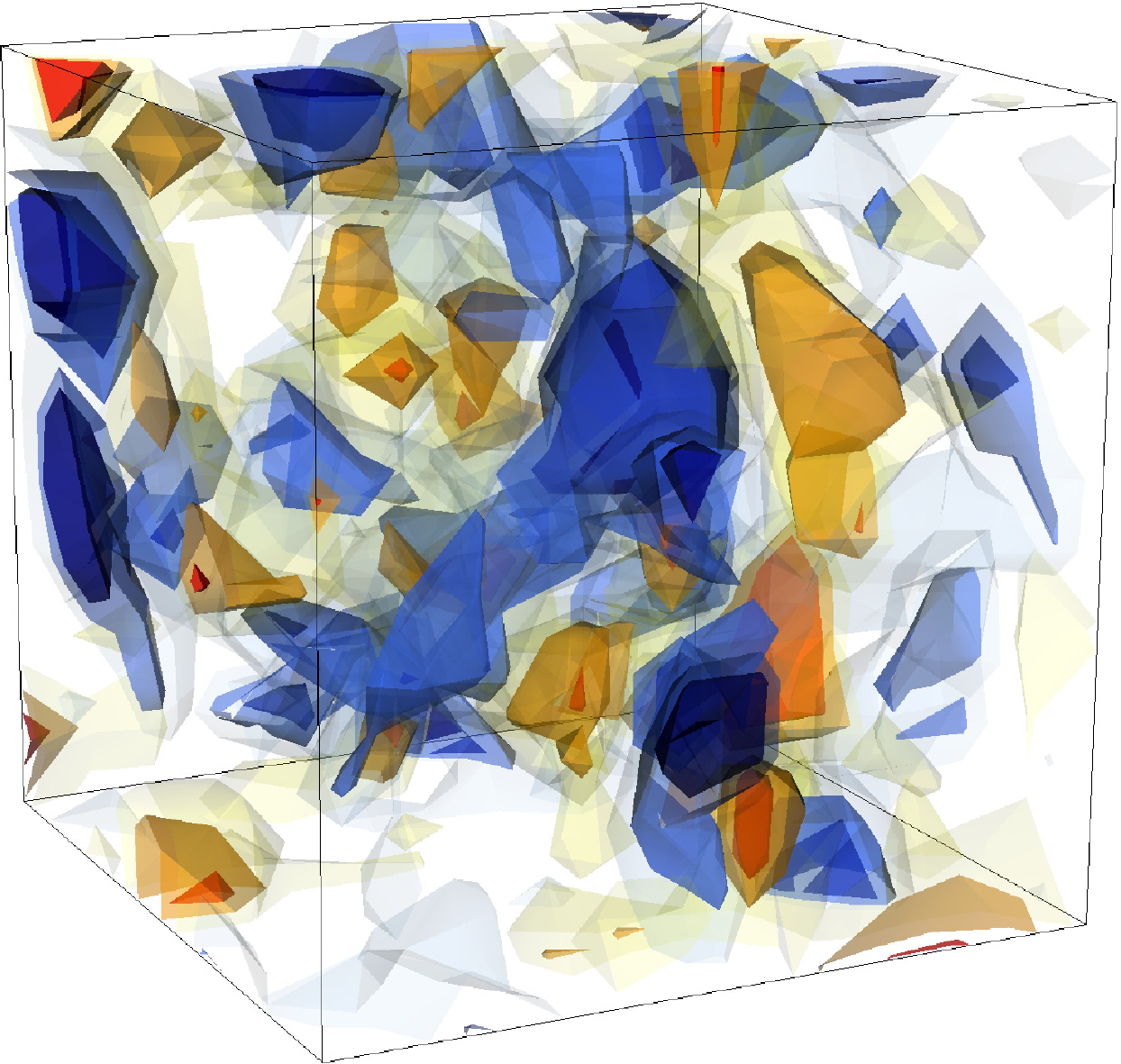}
\end{minipage} & 
 \begin{minipage}[r]{0.08\textwidth}
\includegraphics[width=\textwidth]{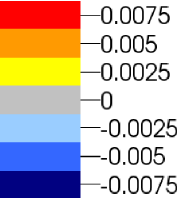}
 \end{minipage}
\end{tabular}
 \caption{Slices of the topological charge density for dynamical overlap ($m_{\pi}=600\,\MeV$), asqtad staggered, dynamical overlap with topology fixing term ($m_{\pi}\approx500\,\MeV$ each) and nonperturbative improved clover (quite heavy $m_{\pi}\approx 1\GeV$) fermions, all with the same lattice spacing  $a=0.12 \fm$. The quenched counterparts of the latter two configurations are also depicted (Iwasaki $\leftrightarrow$ top.fixed overlap and plaquette $\leftrightarrow$ clover). The lattice volume is $12^{3}$ in all cases corresponding to a size of approximately $(1.5 \fm)^{3}$. The color scale is equal in all plots: Blue represents negative and red positive topological charge density.} 
 \label{fig:samespacing}
\end{figure*}

In \fig\ref{fig:samespacing} we show the topological structure of one sample configuration 
for different fermion actions, after 5 steps of improved stout smearing. They all have the same lattice spacing 
of $a=0.12\fm$ and the same physical volume $\text{V}=(1.44 \fm)^{3}$. While this is only a small fraction of the total four-dimensional volume, one can see 
already some of the main properties. Dynamical lattice simulations tend to give larger fluctuations of the topological charge density 
than their quenched counterparts. This was already pointed out in, e.g.,  
Ref.~\cite{Moran2008}. The magnitude of these fluctuations and the individual profiles strongly depend on the action.

\begin{figure*}[t!!]
 \centering
\begin{tabular}{cccccc}
\begin{minipage}[]{0.2\textwidth}
\includegraphics[width=1\textwidth]{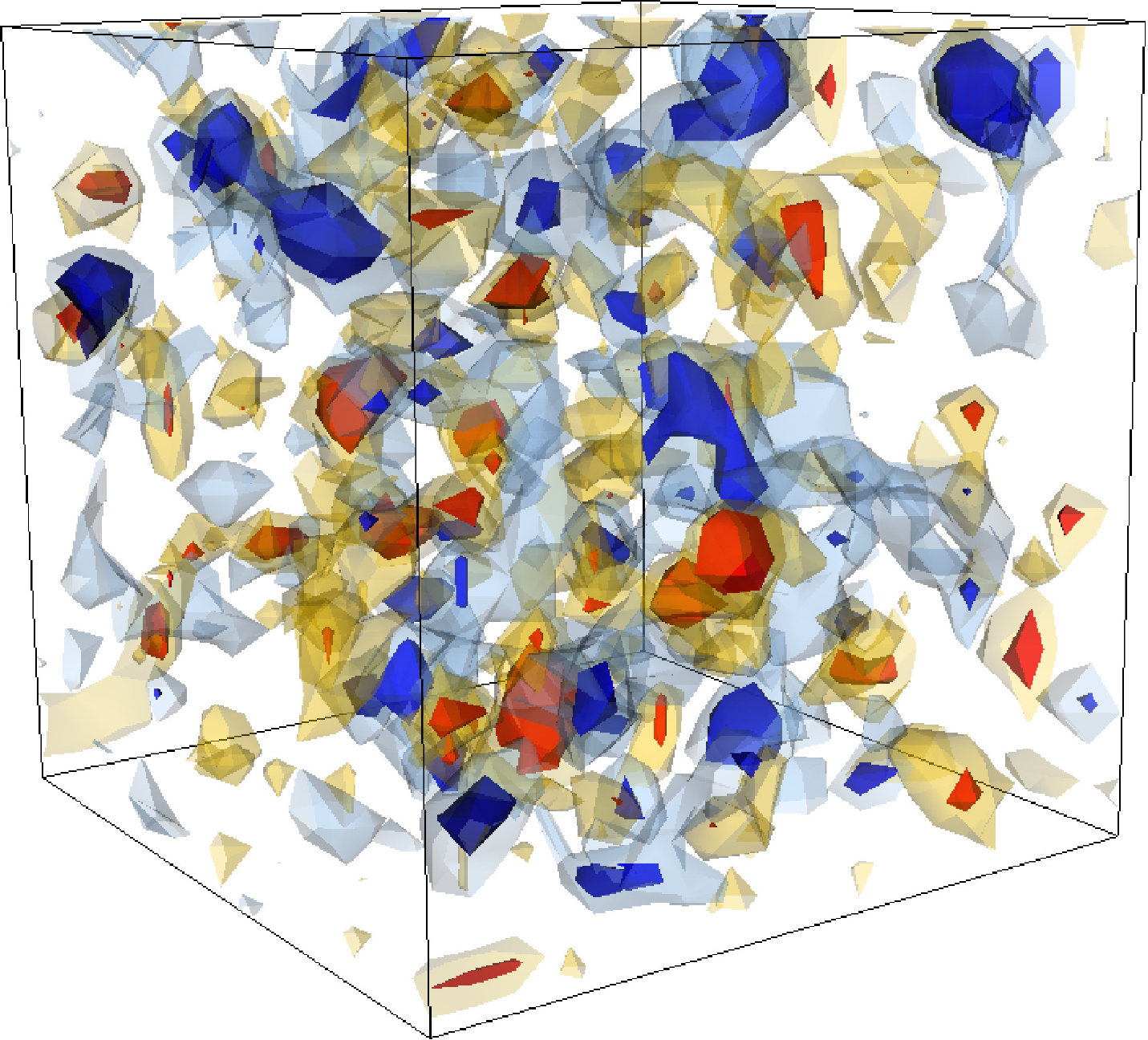}
\end{minipage}  &$\Rightarrow$&\begin{minipage}[]{.2\textwidth}
\includegraphics[width=1\textwidth]{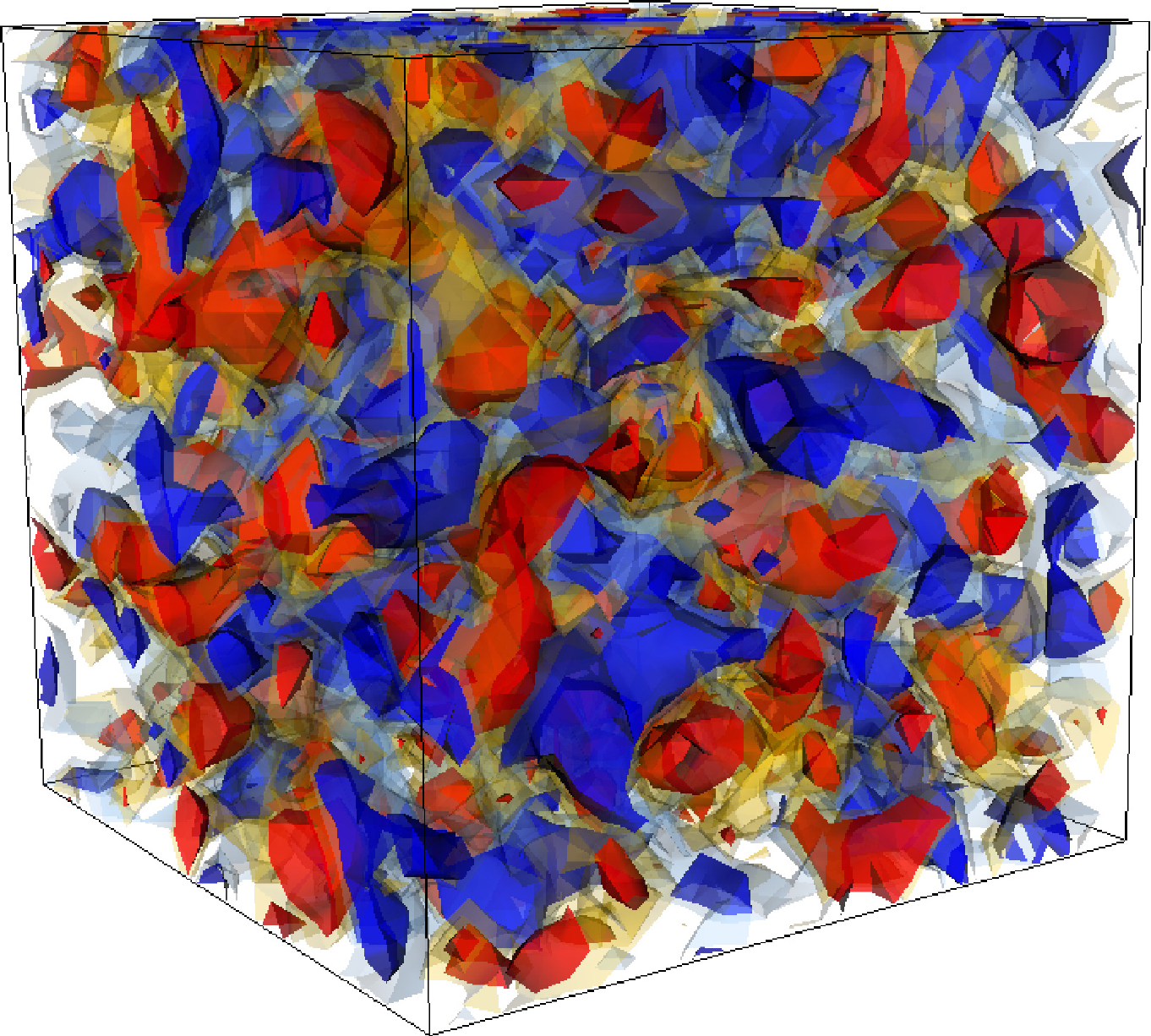}
\end{minipage} &$\Rightarrow$&\begin{minipage}[]{0.2\textwidth}
\includegraphics[width=1\textwidth]{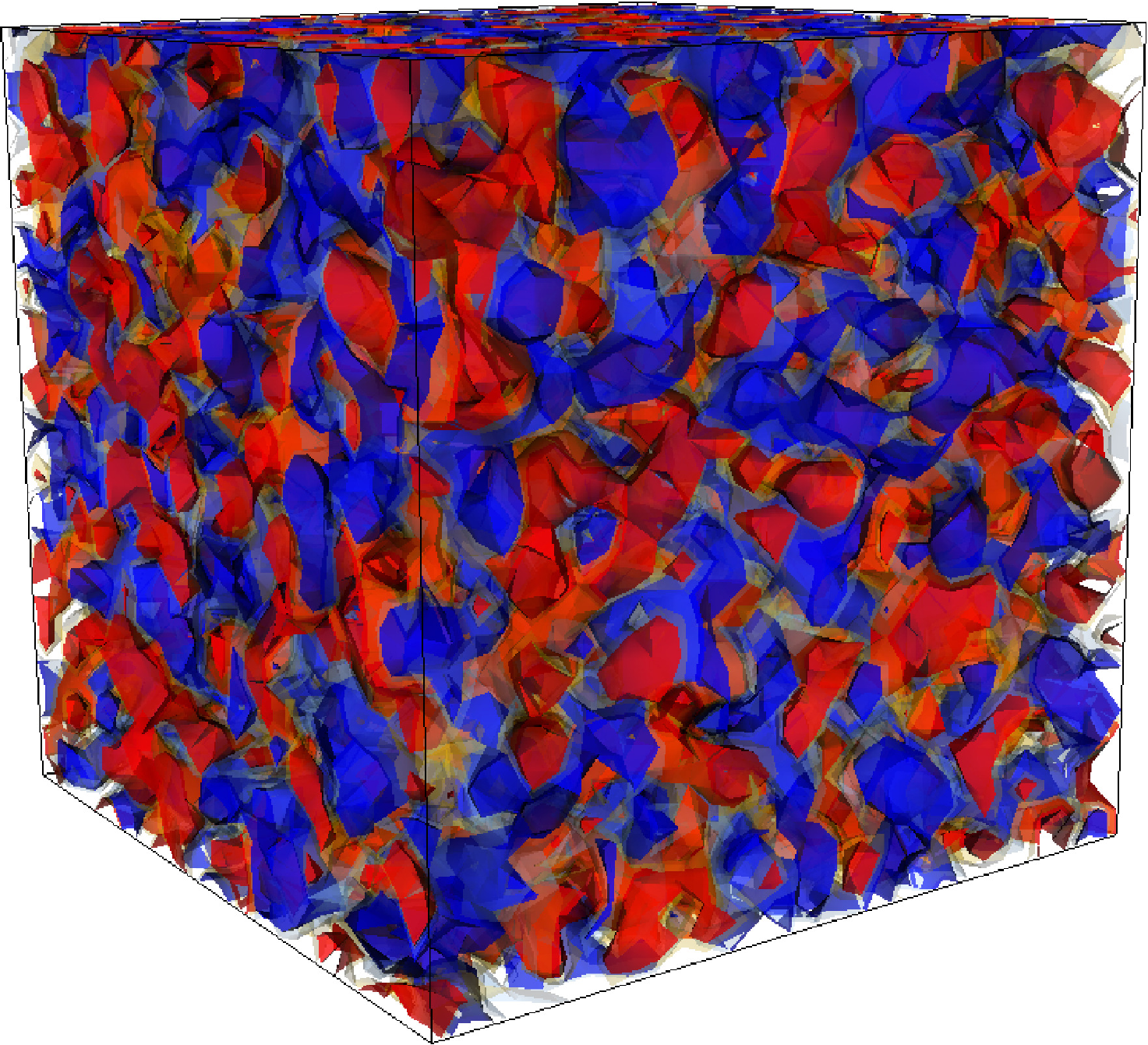}
\end{minipage}& 
\begin{minipage}[r]{0.085\textwidth}
\bigskip
\includegraphics[width=\textwidth]{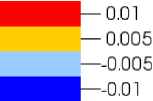}
 \end{minipage} \\ \smallskip\\
\textbf{a=0.1fm}& &\textbf{a=0.079fm} &&\textbf{a=0.063fm} &   
\end{tabular}
 \caption{3d slice of the topological charge density for twisted mass fermions at fixed spatial volume $V= (1.9 \fm)^{3}$ after 5 steps of improved stout smearing. The color scale is equal in all three slices. We see an increasing laminar structure \cite{Horvath:2003yj} for finer lattices.}
 \label{fig:finerandfiner}
\end{figure*} 

A direct comparison of the profiles from different actions is only possible if the lattice spacings are equal 
because the densities $q(x)$ scale with $a^{-4}$. The strong dependence on the lattice spacing is visualized in \fig\ref{fig:finerandfiner}. 
Two main effects towards the continuum limit are clearly visible. First, the structure becomes more and more fine grained and second the strength of the density in physical units increases. To quantify this observation we measure the two-point correlation function of the topological charge density after 5 steps of improved stout smearing. 

\begin{figure}[b!!]
\centering
\includegraphics[width=0.7\textwidth]{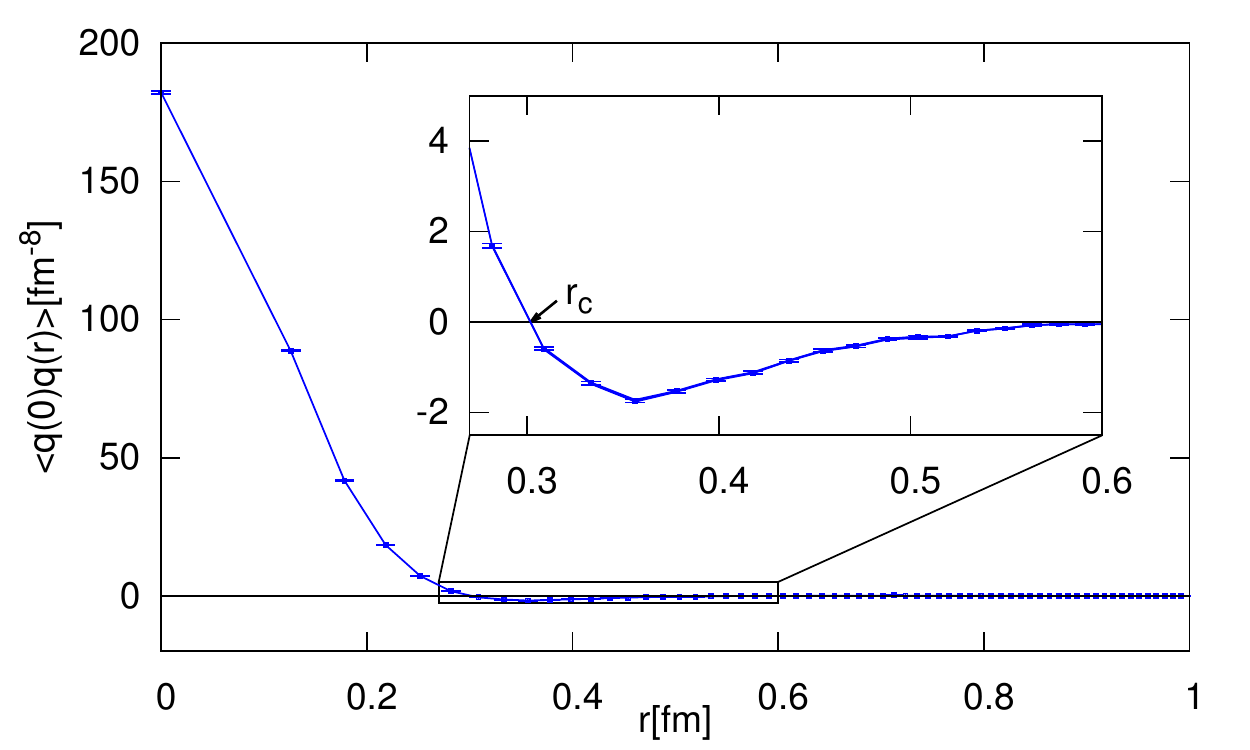}
 \caption{Two-point function of the topological charge density after 5 improved stout smearing steps for an ensemble of dynamical overlap configurations.}
 \label{fig:q0qxexample}
\end{figure}

In \fig\ref{fig:q0qxexample} we show this two-point function for a dynamical overlap ensemble. The correlator has the same characteristic behavior for all fermion actions. At small distance the correlator develops a positive core of radius $r_{\rm c}$, for large distance it is compatible with zero and in between it is slightly negative. 

In the continuum, however, this correlator is negative for any finite distance $|x-y|>0$ and has a positive contact term at zero distance which overcompensates the negative contribution to the space-time integral in Eq.~(\ref{eq:topsuceptibility}) to yield a positive topological susceptibility. This follows from reflection-positivity of the theory and the pseudo-scalar nature of the topological charge density \cite{Seiler}. Hence, the core size has to vanish and the zero distance correlator has to diverge in the continuum limit.
 
 \begin{figure*}[t!]
 \centering
 \includegraphics[width=0.9\textwidth]{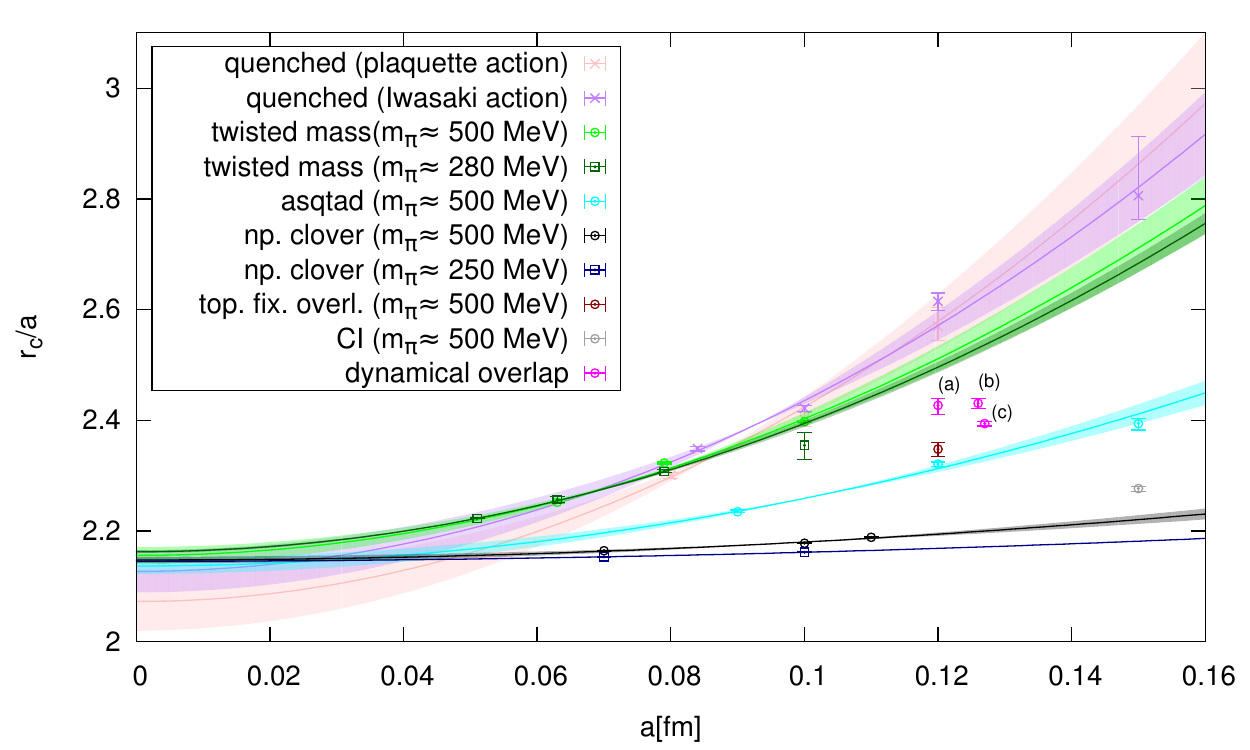}
 \caption{Zero of the 2-point function of the topological charge density. The errors -- which are partly to small to be seen -- result from the interpolation between the points of the correlator. We included fits and error bands for twisted mass, asqtad and clover action as well as for two different quenched simulations. There are three different points from the overlap ensembles at different masses: (a) $600\,\MeV$, (b) $560\,\MeV$ and (c) $510\,\MeV$.}
 \label{fig:zeroinlatticeunits}
\end{figure*}

We show the core size $r_{c}$ for different actions and lattice spacings in \fig\ref{fig:zeroinlatticeunits}. For the twisted mass, the nonperturbative clover and the asqtad staggered fermion action we have enough data to extrapolate to zero lattice spacing. 
We expect discretization errors which are independent of the action used and errors related to these actions. As all actions are $\BigO(a)$-improved
the latter should only set in at  $\BigO(a^2)$. Therefore, we fit the data with a function of the form 
\begin{equation}
    r_{c}/a=C+B\cdot a^{2}\,.
\end{equation}
We find that the coefficients $C$ almost coincide for all actions and are compatible with $C=2.15$ and that the $B$'s differ \cite{Bruckmann2011b}. If we compare the curves for twisted mass and clover fermions at $m_{\pi}\approx 500\, \MeV$ and  $m_{\pi}\approx 280 (250)\,\MeV$, 
we find a relatively weak dependence on the mass. \fig\ref{fig:zeroinlatticeunits} also includes the results for dynamical overlap fermions with and without topology fixing term at similar lattice spacings. The data points for dynamical overlap configurations are not on one line, because they belong to 
different pion masses and, hence, different fit curves. 
 
We have also included two quenched results for comparison. The extrapolations yield a consistent value for the constant $C$ with a slightly larger error than the dynamical configurations. The quenched results  are in accordance with the results of Horvath \etal~\cite{Horvath:2003yj}. They used the fermionic definition of the topological charge including all Dirac modes and find a core size $r_{c}\approx 2a$ 
for the Iwasaki gauge action. 

 \begin{figure}[t]
\centering
 \includegraphics[width=0.7\textwidth]{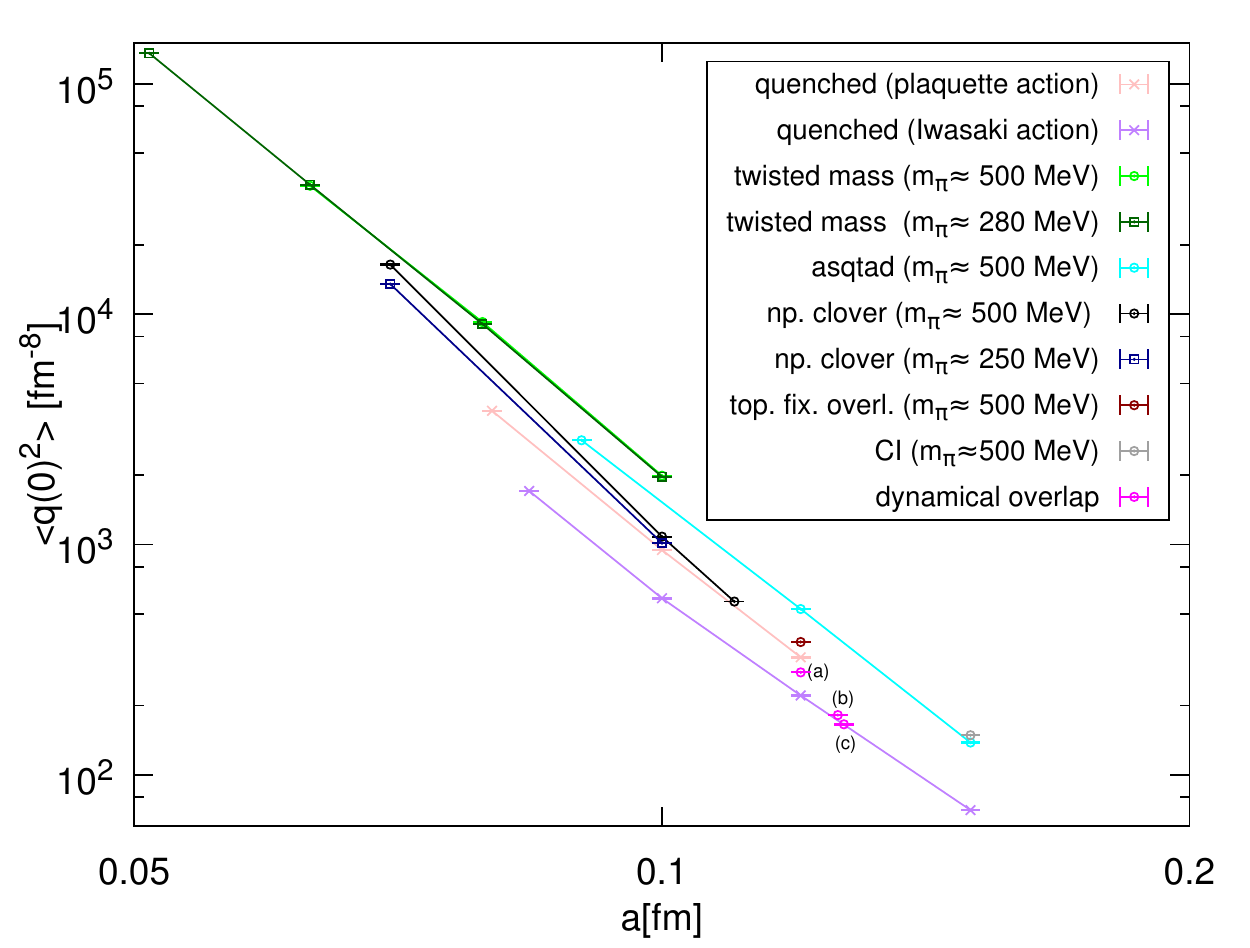}\\
 \caption{Maxima of the two-point function of the topological charge density (overlap results are labeled as: (a) $600\,\MeV$, (b) $560\,\MeV$ and (c) $510\,\MeV$). Note that the twisted mass results for $m_{\pi}=280\, \MeV$ and $500\,\MeV$ fall on top of one another.}
 \label{fig:maxima} 
\end{figure}

The mean-square value of the topological charge density $\langle q(0)^{2}\rangle$ has to diverge 
in the continuum limit. \fig\ref{fig:maxima} shows a double logarithmic plot of $\langle q(0)^{2}\rangle$ 
in physical units versus the lattice spacing. The linear behavior indicates a power-like divergence for $a\to0$ for all actions.

This observable seems to be very sensitive to the action used. The curves do neither fall on top of each other nor do they have the same slope. It is also important to notice, that the contact term for dynamical overlap configurations at $a=0.12\fm$ lies between the two quenched simulations (plaquette and Iwasaki). Thus we can conclude, that dynamical fermions generate indeed larger fluctuations in the vacuum at finite lattice spacing, as argued in \cite{Moran2008}, but also that the differences between different actions are as large as those between quenched and dynamical simulations.

\section{Summary} 
We have investigated the topological charge density for state-of-the-art lattice actions with dynamical fermions, including new dynamical overlap simulations. The radius of the positive topological charge density core approaches zero in the continuum limit with the same slope $C$ for all actions. At finite lattice spacing the spread is below $10\%$ and even quenched simulations do not produce markedly different results.
Also, simulations with exact overlap fermions give results which are 
quite similar to those obtained with topology fixed overlap fermions. The differences between quenched and dynamical simulations are not larger than those between different dynamical fermion actions. 

In contrast, the effects for $\langle q(0)^{2}\rangle$ are large but unsystematic. These results are very sensitive to the lattice spacing $a$,
implying that one should be very careful not to jump to conclusions when comparing topological properties of different 
configurations.

If we use our dynamical overlap results as benchmark for the quality of the other actions with respect to chirality we have to conclude that all of them are reasonable successful and none of them seems to be clearly superior. The differences between results for dynamical overlap fermions and topology fixed overlap fermions are especially small, as one might have expected. For more details we refer to Ref.~\cite{Bruckmann2011b}. \\

We would like to thank Gunnar Bali and Ernst-Michael Ilgenfritz for insightful discussions. Furthermore, we want to thank C.B. Lang, M. Limmer and D. Mohler who generated the chirally improved configurations which were used in this Letter. FB has been supported by DFG BR 2872/4-2 and FG by SFB TR-55 as well as by a grant of the ''Byerische Elitef\"orderung``. We thank NIC, J\"ulich and LRZ, Garching for providing the computer time used to generate the dynamical overlap fermion and CI configurations, respectively. This work was also supported by STRONGnet.
 
\bibliographystyle{myh-physrev}
\bibliography{overlaptopology}
\end{document}